# Thermomechanical Behavior of the HL-LHC 11 Tesla Nb₃Sn Magnet Coil Constituents during Reaction Heat Treatment

C. Scheuerlein[1], F. Lackner[1], F. Savary[1], B. Rehmer[2], M. Finn[2], C. Meyer[2]

[1] *European Organization for Nuclear Research (CERN), Geneva, Switzerland*
[2] *Federal Institute for Materials Research and Testing (BAM), Berlin, Germany*

*Abstract*—The knowledge of the temperature induced changes of the superconductor volume, and of the thermomechanical behavior of the different coil and tooling materials is required for predicting the coil geometry and the stress distribution in the coil after the Nb₃Sn reaction heat treatment. In the present study we have measured the Young's and shear moduli of the HL-LHC 11 T Nb₃Sn dipole magnet coil and reaction tool constituents during *in situ* heat cycles with the dynamic resonance method. The thermal expansion behaviors of the coil components and of a free standing Nb₃Sn wire were compared based on dilation experiments.

## I. INTRODUCTION

THE superconducting magnets [1] for the LHC High Luminosity upgrade (HL-LHC) [2] are built using Nb₃Sn technology. The use of the brittle Nb₃Sn implies a reaction heat treatment (HT) when the conductor is at its final shape after coil winding. During this HT Nb₃Sn conductor volume changes take place, and at the end of the HT a permanent coil cross section and length change is commonly observed. While the Nb₃Sn coil cross section generally increases, in some magnet types a coil length shrinkage is observed [3], and in others a length increase occurs.

In order to predict the coil volume changes by finite element modelling, the thermal expansion behavior of the different coil and reaction tool constituents, as well as the temperature dependence of their elastic and plastic materials properties need to be known.

It is obvious from the comparison of the room temperature (RT) stress-strain curves of the 11 T dipole magnet constituents (Fig. 1) that the different coil materials mechanical properties are varying strongly. More details about the RT stress strain measurements can be found in [4].

In this article we describe the temperature dependence of these materials properties, based on dynamic Young's and shear modulus measurements during *in situ* HT up to 700 °C, and we have estimated the Poisson's ratio temperature evolution from *E* and *G*.

C. Scheuerlein, F. Lackner and F. Savary are with CERN, CH-1211 Geneva 23, Switzerland, (corresponding author phone: ++41 (0)22 767 8802, e-mail: Christian.Scheuerlein@cern.ch).
B. Rehmer, M. Finn and C. Meyer are with Federal Institute for Materials Research and Testing (BAM), Berlin, Germany.

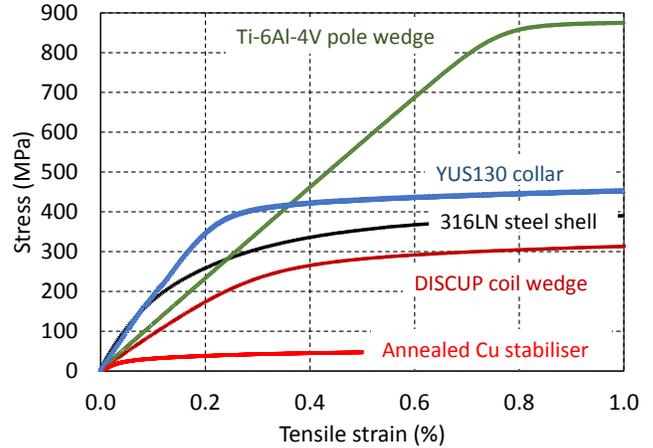

Fig 1: Comparison of the RT engineering stress-strain curves of annealed superconducting wire Cu stabilizer, DISCUP C3/30 coil wedge, Ti6Al4V pole wedge, YUS130 collar, and stainless steel 316LN (magnet shell and reaction fixture) [4].

For the calculation of elastic properties from resonance frequencies the thermal expansion coefficients need to be known. The thermal expansion mismatch of the conductor and the other coil materials leads also to stresses in the coil that influence the coil geometry at the end of the heat cycle. Therefore, the thermal expansion coefficients have been determined by dilation experiments.

## II. EXPERIMENTAL

### A Resonance Testing

Dynamic Young´s and shear modulus measurements have been performed by the resonance technique [5]. The Young´s and the shear moduli were calculated from the specimen dimensions, density, and resonance frequencies for bending and for torsion, respectively. Dynamic measurements were performed *in situ* during HT in a furnace in inert gas. As an example, the temperature dependent shift of the first bending oscillation of the 316LN specimen is shown in Fig.2.

….

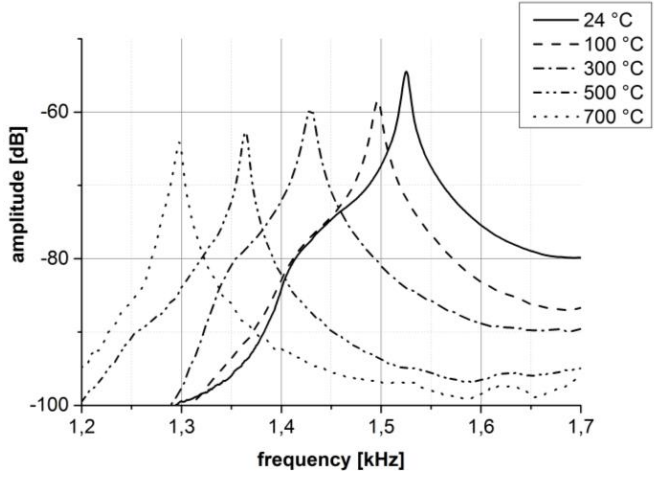

Fig 2: Resonance peak of the first bending oscillation of the 316LN specimen at different temperatures.

Rectangular specimens with the dimensions 3 mm×9 mm×100 mm have been used. For more information about the dynamic measurements and the calculation of Young's modulus from the fundamental resonance frequency of the beam in bending see reference [4]. For long thin beams with rectangular cross section the shear modulus (G) is calculated from the mass ($m$), length ($L$), width ($b$) and thickness ($t$) of the beam, and its fundamental resonance frequency of the beam in torsion ($f_t$),

$$G = \frac{4Lmf_t^2}{bt}\left(\frac{B}{1+A}\right) \quad (1)$$

where

$$B = \frac{\frac{b}{t}+\frac{t}{b}}{4\frac{t}{b}-2.52\left(\frac{t}{b}\right)^2+0.21\left(\frac{t}{b}\right)^6} \quad (2)$$

The term A can be ignored if the ratio 1<$b/t$<2. For isotropic materials Poisson´s ratio can be calculated from the Young's modulus ($E$) and $G$ according to Equation (3):

$$\mu = \frac{E}{2G}-1 \quad (3)$$

*B Direct Poisson's ratio measurement*

DISCUP C30/3 and Ti6Al4V Poisson's ratio measurements have been performed in tension using Ø=8 mm samples that have been extracted from 11 T dipole wedges by wire electro discharge machining (Fig. 3).

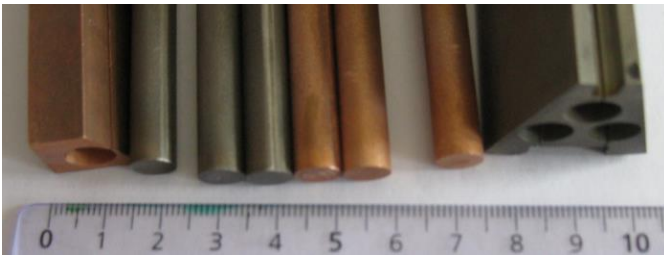

Fig. 3: Samples for Poisson's ratio measurements.

The set-up for Poisson's ratio measurements consists of two extensometers for determination of longitudinal and transversal strain during tensil loading (Fig. 4). The Poisson's ratio is calculated as the quotient of the thickness change by the change in length for elastic deformation according to:

$$\mu = \frac{\Delta d/d_0}{\Delta l/l_0} \quad (4)$$

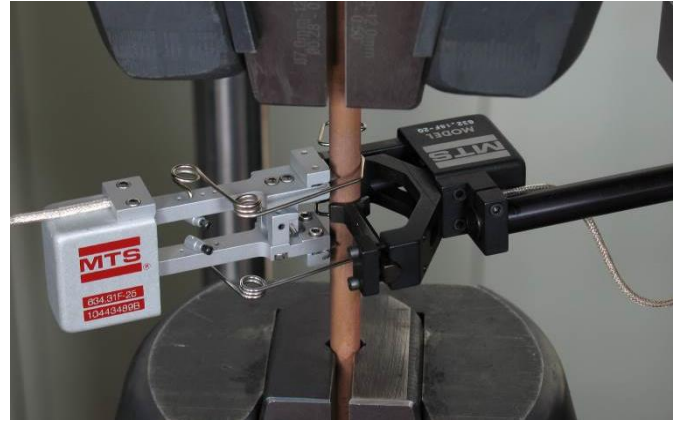

Fig. 4: Set-up for Poisson's ratio measurement with one extensometer for axial and one for radial strain measurement.

*C Thermal expansion measurements*

The measurement of the temperature dependent expansion was performed in the temperature range between -140 °C and 700 °C with an electronical highly vacuum-tight horizontal push rod dilatometer by Netzsch (DIL402c). The test procedure and test analyses was carried out in accordance with [6]. The specimens (4 mm×4 mm×25 mm) were heated in argon atmosphere with a heating rate of 5 K/min. During heating-up the change in length $\Delta l/l_0$ was measured. The mean technical coefficient of thermal expansion $\alpha_T(T_0;T_1)$ is calculated according to:

$$\alpha_T(T_0;T_1) = \frac{\Delta l}{l_0 \times \Delta T} \quad (5)$$

with $T_0$=20 °C. The values shown in Fig. 7 are average values of two individual measurements.

Thermal expansion measurement of an $Nb_3Sn$ and a Cu wire in axial direction have been performed with a Netzsch 402E high temperature dilatometer. $Nb_3Sn$ wire samples are not well suited for dilatometry measurements, since they are always slightly bended, have internal stresses and the wire ends are modified during the measurements by the phase changes occurring inside the wire and the Sn melting in particular. The absolute $Nb_3Sn$ wire dilation results may therefore not be highly accurate, but they show trends of the wire length changes during the reaction HT and subsequent cooling.

*D 11 T coil and reaction fixture materials*

A 11 T coil cross section with its 6 conductor blocks and 5 coil wedges is presented in Fig. 5. The 11 T dipole $Nb_3Sn$ coils are wound from non-reacted $Nb_3Sn$ keystoned Rutherford type cable [7,8] with its 0.15 mm-thick insulation made of a Mica tape and S2/E-glass fibre. In order to keep the coil shape during subsequent manipulation, a binder CTD-1202 from Composite Technology Development is added.

....

The brittle Nb₃Sn is formed during a reaction HT with a peak temperature of typically 650 °C. In order to improve the electromechanical coil properties, the void space in the reacted coils is filled with epoxy resin CTD-101K from Composite Technology Development [9].

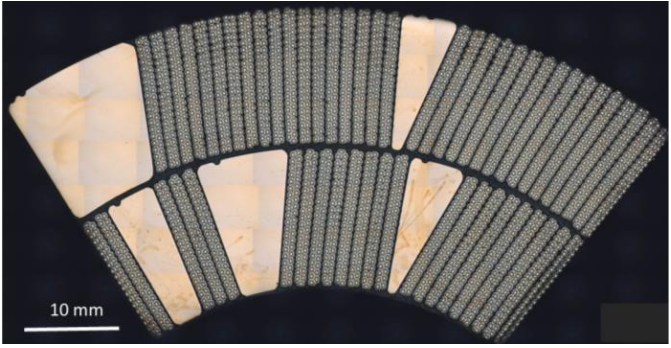

Fig. 5: Metallographic cross section of 11 T dipole coil. Courtesy of Mickaël Meyer, CERN.

In Fig. 6 a reacted 11 T dipole coil pole end is shown inside the coil reaction fixture, which is made of stainless steel 316LN. During the HT a 316LN pole wedge is used, which is replaced after the HT by a Ti6Al4V pole wedge. It can be seen that after the reaction HT a gap remains between the Nb₃Sn coil and the 316LN wedge, indicating that a permanent coil length increase occurred during the HT.

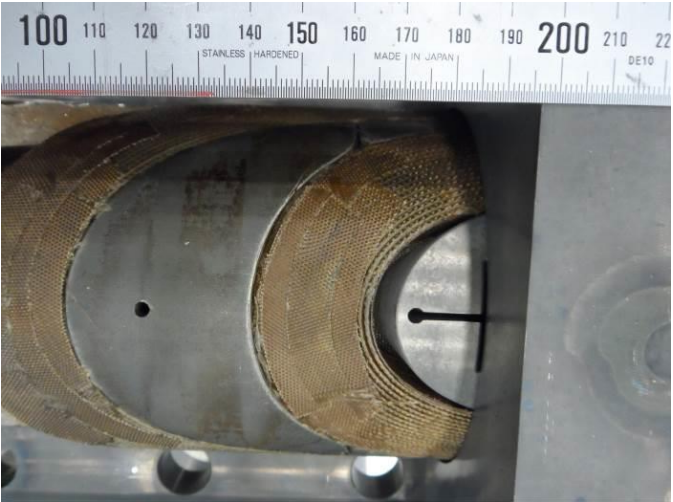

Fig. 6: Reacted 11 T dipole coil pole end inside the 316LN reaction fixture.

The mechanical coil block properties are usually roughly estimated from transverse compression tests of so-called ten-stack samples, which show that the composite coils exhibit strongly non-linear stress-strain behavior and a relatively low stiffness [10]. The different superconductor materials are strongly textured [11] and exhibit anisotropic materials properties [12].

The 11 T dipole coils are mechanically loaded through the removable pole wedges made of Ti6Al4V. The coil wedges between the conductor blocks are made of aluminum oxide dispersion strengthened ODS Copper known under the tradename "CEP DISCUP® C3/30". Cu dispersion strengthening with ultrafine Al₂O₃ particles maintains high strength during the Nb₃Sn coil 650 °C-50 h HT to which the wedges are submitted. CEP DISCUP® C3/30 contains about 0.6 wt.% Al, and has an ultra-fine structure with a texture in the direction of extrusion [13,14]. All test specimen have been cut in the extrusion and drawing direction of the wedges.

Dilatation measurements during reaction HT have been performed in axial direction of an about 10 mm long free standing Nb₃Sn wire. The RRP type wire fabricated by OI-ST (billet #7419) contains 54 Nb–Ta alloy filament bundles, each surrounded by distributed diffusion barriers. It has a diameter of 0.803 mm (before reaction) and 0.823 mm (after reaction). The overall length change measured at RT before and after reaction HT of a free standing RRP #7419 wire is with -0.07% much smaller than the wire cross section increase of +4.9% [15].

### III. RESULTS

*A Temperature dependence of the Young's and shear moduli*

The resonance test results of the DISCUP, Ti6AlV4 and steel 316LN Young's moduli and shear moduli as a function of temperature are presented in Fig. 7 and Fig. 8, respectively.

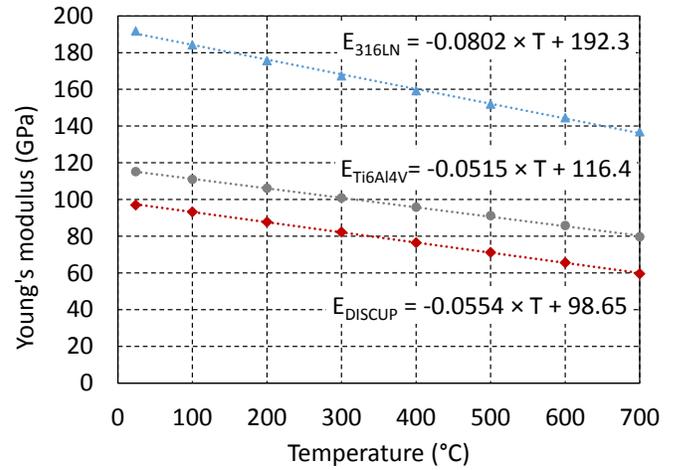

Fig. 7: Temperature dependence of the DISCUP C3/30 wedge, Ti6Al4V pole wedge and 316LN stainless steel Young's moduli and linear fits.

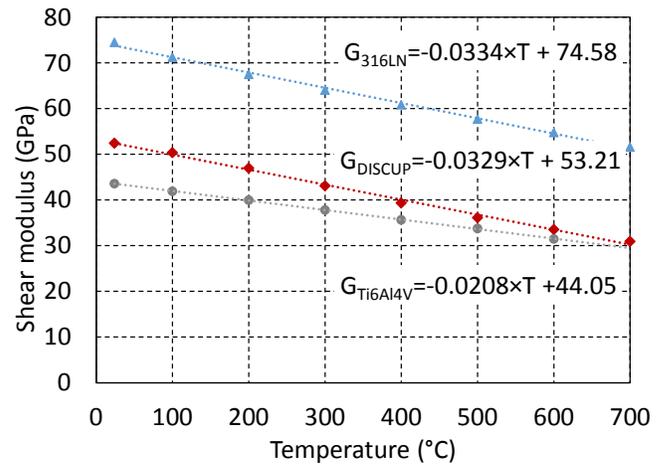

Fig. 8: Shear moduli of the DISCUP C3/30 wedge, the Ti6Al4V pole wedge and 316LN steel as a function of temperature.

The temperature dependence of the Poisson's ratio calculated from $E$ and $G$ according to Equation 3 is presented in Fig. 9 for steel 316LN and Ti6Al4V. For the DISCUP wedges no reasonable Poisson's ratio values could be

…

calculated with the simple relation, presumably because the highly textured DISCUP wedges exhibit strongly anisotropic materials properties.

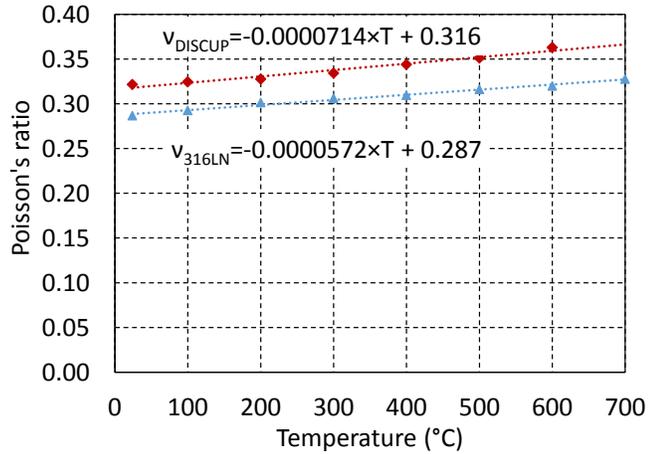

Fig. 9: Temperature dependence of the Ti6Al4V pole wedge and 316LN steel Poisson's ratio calculated from Young's and shear moduli.

The linear fits of the temperature dependent Young's and shear moduli, as well as the Poisson's ratios calculated from E and G are summarized in Table I. By extrapolating the linear fits, rough estimates of the 4.2 K Young's moduli of 216 GPa, 130 GPa and 113 GPa are obtained for 316LN, Ti6Al4V and DISCUP C3/30, respectively.

TABLE I
Young's, Shear Moduli and Poisson's ratio as a function of temperature (T) in °C in the temperature range 20 °C to 700 °C.

| Material | Young's modulus (GPa) | Shear modulus (GPa) | Poisson's ratio |
|---|---|---|---|
| 316LN_L | $-0.080 \times T + 192$ | $-0.033 \times T + 74.6$ | $0.000057 \times T + 0.29$ |
| Ti6Al4V_L | $-0.052 \times T + 116$ | $-0.021 \times T + 44.1$ | $0.000071 \times T + 0.32$ |
| DISCUP_L | $-0.055 \times T + 98.7$ | $-0.033 \times T + 53.2$ | |

The RT results of 316LN are slightly lower than the values of E=193 GPa, G=77 GPa and ν=0.25 reported in [16]. Also the Ti6Al4V RT results match reasonably well with the values of E=113.8, G=44.0 reported in [17].

The DISCUP RT Youngs modulus is lower and the shear modulus higher than the corresponding values E=110 GPa, G=46 GPa reported for annealed Cu [18]. The Poisson's ratio calculation from $E$ and $G$ according to Equation 3 yielded negative values, showing that the simple relation between $E$, $G$ and ν does not apply in the case of the strongly textured and anisotropic DISCUP wedges.

*B Direct DISCUP and Ti6Al4V Poisson's ratio measurement*

Poisson's ratios of the wedge materials DISCUP C30 and Ti6Al4V determined with the static set-up shown in Fig. 4 are $\nu_{DISCUP-static}$=0.43±0.02 and $\nu_{Ti6Al4V-static}$=0.32±0.03, respectively (average of three measurements ± one standard deviation).

The static Ti6Al4V Poisson's ratio result is in good agreement with the RT value $\nu_{Ti6Al4V-dynamic}$=0.322 calculated from the the dynamic E and G results, and it matches within the experimental uncertainties with the value of 0.342 reported elsewhere [17].

The static DISCUP C30 Poisson's ratio is significantly higher than the values reported elsewhere of ν=0.343 [18] and ν=0.364 for annealed and for cold-drawn Cu [19], respectively.

*C Thermal expansion of $Nb_3Sn$ coil constituents*

In Fig. 10 the relative length changes of the coil constituents DISCUP C3/30, Ti6Al4V and the tooling material 316LN are compared with those of an initially unreatced $Nb_3Sn$ RRP type wire during the coil reaction HT. For comparison the axial length change of an initially hard drawn Cu wire, and of the Nb thermal expansion reported in [20] are shown as well.

The thermal expansion coefficients of Ti6Al4V and Nb are about 40% and 60% smaller than those of the 316LN, DISCUP and the Cu wire, which exhibit similar thermal expansion behavior.

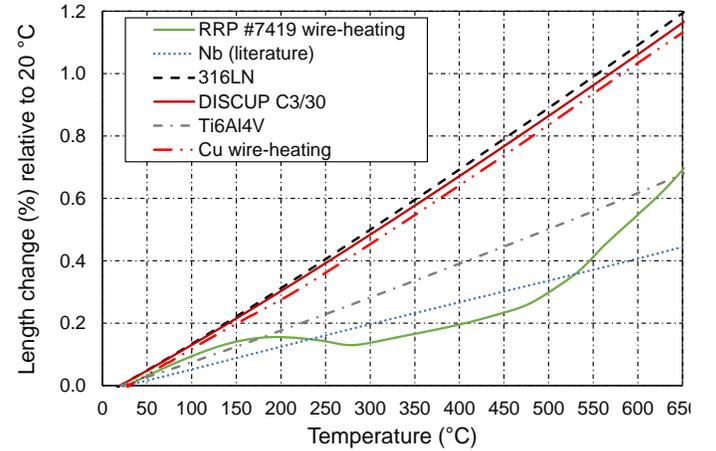

Fig. 10: Comparison of RRP #7419 $Nb_3Sn$ wire axial length change during first heating with that of DISCUP C3/30, Ti6Al4V, and 316LN. The relative length changes of a Cu wire and the Nb thermal expansion from reference [20] are shown for comparison.

It can be seen in Fig. 10 that the length changes of a free standing $Nb_3Sn$ RRP type wire during first heating are rather complicated. Up to about 180 °C the $Nb_3Sn$ wire length increase is inbetween that of Cu and Nb, indicating that the cold-drawn Cu stabiliser can carry a certain axial stress that is caused by the thermal expansion mismatch of the Cu stabiliser and the Nb filaments.

Upon further heating to approximately 270 °C the wire shrinks, presumably due to relaxation of the Nb filaments when the initialy cold-drawn Cu stabilsier is annealed and looses its mechanical strength [21,22]. During subsequent heating to about 480 °C the wire length increases with a slope that is similar to the thermal expansion of Nb. A similar trend is observed in the Nb axial lattice parameter variations reported for the same wire in [23], and in dilation results obtained during first heating of another internal tin type $Nb_3Sn$ wire [24].

Upon further heating a comparatively strong axial wire length increase is observed. The strong increase of thermal expansion occurs at similar temperature as the melting of the brittle intermetallic $Cu_6Sn_5$, formation of $(Nb_{0.75}Cu_{0.25})Sn_2$ [25] and other Nb containing intermetallic phases [15]. The onset of Nb$_3$Sn formation in this RRP wire is at about 550 °C. The wire length increase continues during isothermal heating at the processing peak temperature (not shown in Fig. 10).

The relative length changes of the same materials during cooling from the reaction peak temperature of 650 °C are shown in Fig. 12. During cooling the shrinkage of the reacted Nb$_3$Sn wire is similar to that of Nb$_3$Sn bulk [20]. This can be explained by the fact that the Nb barrier cross section remaining in the wire after reaction HT is only a small fraction of the Nb$_3$Sn cross section. Therefore, Nb$_3$Sn dominates the thermal expansion behaviour of the composite.

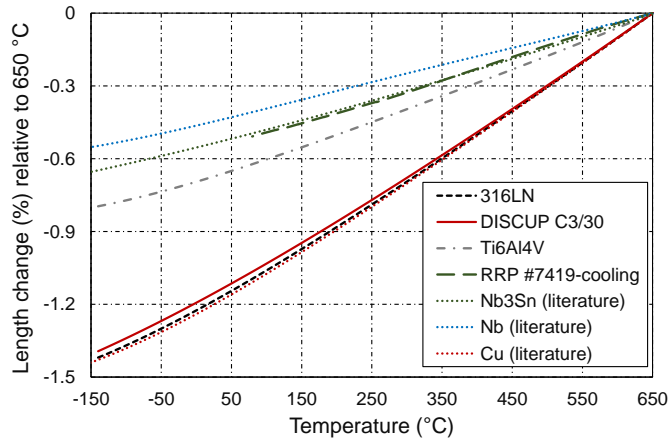

Fig. 12: Relative length change of DISCUP C30/3, Ti6Al4V, 316LN and the reacted Nb$_3$Sn wire #7419 during cool down from 650 °C. The thermal expansions of Cu, Nb and Nb$_3$Sn bulk [20] are shown for comparison.

## IV. DISCUSSION AND CONCLUSION

The coil volume changes that occur during its reaction HT are one of the main challenges to be adressed in the manufacturing of Nb$_3$Sn accelerator magnets. The coil geometry and stresses in the coil are influenced by a variaty of materials properties, including the friction coefficients between the different components [26], cabling and coil winding parameters, and by the tooling materials and geometry. Here we report the thermal expansion and the temperature dependent elastic properties of the coil and tool materials.

The unusual thermal expansion behavior during the first HT after the Nb$_3$Sn wire cold drawing process, and its thermal expansion mismatch with the other coil materials like the DISCUP coil wedges, Ti6Al4V pole wedges and the 316LN stainless steel tooling lead to stress formation in the different components during the thermal cycles. At the end of the Nb$_3$Sn formation HT the wire exhibitis a thermal expansion behavior resembling that of Nb$_3$Sn bulk.

The absolute thermal expansion of a free standing wire differs from that of cables [15],[27],[28],[3], but we believe that the overall length change behaviour during first heating and subsequent cool down is representative for the same wire in Rutherford cables. The thermal expansion behavior of the corresponding Nb$_3$Sn Rutherford cables is subject of further studies.

To derive the mechanical properties needed for the simulation of the geometrical coil behavior over a temperature interval of almost 1000 K by destructive stress-strain measurements is a very laborious procedure. Furthermore, most of the coil materials exhibit non-linear stress strain behavior even at low strains (Fig. 1), which can cause large uncertainties in the E-moduli derived from stress-strain data. Non-destructive dynamic tests are an elegant way for determining the Young's and shear moduli of the coil and reaction fixture materials over the entire temperature range of the Nb$_3$Sn processing HT.

For homogeneous isotropic materials simple relations exist between E-modulus, shear modulus, bulk modulus and Poisson's ratio. These relations do not apply to materials with anisotropic materials, like the DISCUP wedges. For the 316LN and Ti6Al4V a reasonable agreement is found between the Poisson's ratio calculated from the resonance frequencies and literature values.

ACKNOWLEDGEMENTS

We are grateful to the CERN central workshop team for the preparation of the test specimen.

This work was supported by the European Commission under the FP7 project HiLumi LHC under Grant GA 284404, co-funded by the DoE, USA and KEK, Japan.

....